\documentclass[APA,STIX1COL]{WileyNJD-v2}
\usepackage{hyperref}
\usepackage{graphicx,wrapfig,lipsum}
\usepackage{graphicx,calc}
\usepackage{ulem}
 \usepackage{lineno}
  \usepackage{amsmath}
\usepackage{bbm}
\usepackage{mathrsfs,amsfonts,dsfont} 
\usepackage{tikz}
\usepackage{multirow}
\usepackage{amsfonts}
\usepackage{lscape}
 \usepackage{dsfont}
 \usepackage{graphicx}
\usepackage{wrapfig}
\usepackage{tikz}
\usepackage{pgfplots}
\usetikzlibrary{arrows,positioning,shapes.geometric}
\usetikzlibrary{matrix,calc,shapes}
\usetikzlibrary{shapes,arrows}
\usepackage{amsmath,bm,times}
\usepackage[T1]{fontenc}
\usepackage{mwe}    
\usepackage{subfig}
\usepackage{multirow}
\usepackage{epsfig}
	\usepackage{color}

	\usepackage{lscape}
\usepackage{mathtools}
\usepackage{setspace}
\usepackage{amssymb}

\newcommand{\vero}[1]{{\color{black}#1}}
\newcommand{\naf}[1]{{\color{black}#1}}
\newcommand{\verob}[1]{{\color{black}#1}}
\newcommand{\veroc}[1]{{\color{black}#1}}

\articletype{Article Type}%

\received{}
\revised{}
\accepted{}

\def\S{{\mathcal S}}
\def\R{{\mathbb R}}
\def\P{{\mathbb P}}
\def\N{{\mathbb N}}
\def\K{{\mathcal K}}

\newlength\myheight
\newlength\mydepth
\settototalheight\myheight{Xygp}
\begin{document}

\title{Recognizing a Spatial  Extreme \vero{dependence structure}: A Deep Learning approach}

\author[1,2]{Manaf AHMED}

\author[2]{V\'eronique MAUME-DESCHAMPS}

\author[2]{Pierre RIBEREAU}

\authormark{Manaf AHMED \textsc{et al}}

\address[1]{\orgdiv{Department of statistics and Informatics}, \orgname{University of Mosul}, \orgaddress{\state{Mosul}, \country{ Iraq}}}

\address[2]{\orgdiv{Institut Camille Jordan, ICJ}, \orgname{Univ Lyon, Université Claude Bernard Lyon 1, CNRS UMR 5208}, \orgaddress{\state{Villeurbanne, F-69622}, \country{France}}}

\corres{Manaf AHMED,\\  \email{manaf.ahmed@uomosul.edu.iq}}


\abstract[Summary]{Understanding the \vero{behaviour of environmental extreme events is crucial for} evaluating \verob{ economic losses, assessing  risks}, health care and 
many other aspects. \vero{In the spatial context, relevant for environmental events, the dependence structure plays a central rule, as it influence joined extreme events and extrapolation on them. 
So that}, recognising or at least \vero{having} preliminary \vero{informations on} patterns of these dependence \vero{structures is a} valuable knowledge for understanding \vero{extreme  events. 
In this study, we address the question of automatic recognition of spatial Asymptotic Dependence (AD) versus Asymptotic independence (AI), using}  Convolutional Neural Network (CNN). \vero{We have 
designed an architecture of Convolutional Neural Network to be an efficient classifier  of the dependence structure}. Upper and lower tail dependence measures \vero{are} used \vero{to train the} 
CNN. \vero{We have tested our methodology on simulated and real data sets: } air temperature data at two meter \vero{over}   Iraq land and Rainfall 
data in the east cost of Australia. }
\maketitle
\section{Introduction}\label{sec:1}
\vero{Understanding extreme environmental events such as heat waves or heavy rains is still challenging. The dependence structure is one important element in this field. } Multivariate extreme value 
theory (MEVT) is \vero{a good} mathematical framework for modelling \vero{the dependence structure of} extremes events \vero{(see for instance \cite{de2007extreme} and 
\cite{embrechts2013modelling})}.  Max-stable \veroc{processes} \vero{consist in an extension} of multivariate extreme value distributions to \vero{spatial processes and provide models for spatial 
extremes (see  \cite{de2006spatial} and \cite{de1984spectral}). These max-stable} processes \vero{are asymptotically dependent (AD). This may not be realistic in practice. \cite{wadsworth2012dependence} 
introduced inverted max-stable processes which are asymptotically independent (AI). Using AD versus AI models has important implications on the extrapolation of joined extreme events (see 
\cite{bortot2000multivariate} and \cite{coles2002models}). So that}, recognising  the class of  dependence structure is \vero{an important task when building models for environmental data. One of 
the main challenges is how to recognise} the dependence structure pattern  of \vero{a spatial} process. Despite various studies dealt with \vero{spatial extremes models, we have not found works 
focused on the question of the automatic determination of AI versus AD for a spatial process. The usual approach is to use Partial Maximum Likelihood estimations after having chosen (from 
exploratory graph studies) a class of models. We propose a first deep learning approach to deal with this question. Many works using deep learning for spatial and spatio-temporal processes have 
been developed but \veroc{none} concerned with AD versus AI (see \cite{wang2019deep})}. 

Artificial Intelligence techniques \vero{have} demonstrated a significant efficiency in many applications, such as environment, risk management, image analysis and many others. \vero{We 
will focus on Convolutional Neural Network (CNN) which have ability in automatic extracting the spatial features 
hierarchically. It has been used e.g. on} spatial dependencies from raw datasets. For instance, \cite{zhu2018wind} \veroc{proposed 
a predictive} deep convolutional neural network to predict the wind speed in \vero{a} spatio-temporal context, \vero{where the spatial dependence  between locations is captured}. 
\cite{liu2016application} developed \verob{a CNN} model to predict \vero{extremes of} climate events, such as tropical cyclones, atmospheric rivers and weather fronts.  \cite{lin2018exploiting} presented 
an  approach to forecast  air quality  (PM2.5 construction) \verob{in a Spatio-temporal} framework. While  \cite{zuo2015convolutional}  improved the power of \vero{recognising  objects in } images by 
learning the spatial dependencies of these regions via CNN. \verob{In the spatial} extreme context,  \cite{yu2016modeling} tried to model the spatial extremes by bridging  the cape between the traditional 
statistical method and Graph methods via decision Trees.
 
\vero{Our } objective is to employ  \veroc{deep learning} concepts \vero{in order to recognise  patterns} of spatial extremes dependence structures \vero{and distinguish between AI and AD.}                
 Upper  and  Lower tail dependence measures \verob{${\chi}$  and ${\bar{\chi}}$} \vero{are used as a summary of the extreme dependence structure.}  These dependence measures 
\vero{have been } introduced by \cite{coles1999dependence} \vero{in order} to quantify the pairwise dependence of \vero{extremes events} between two locations. \vero{Definitions and properties } of 
these measures will be \vero{given} in Section \ref{sec:2}.  The pairwise empirical \verob{versions of these measures are used as a summary dataset}. The CNN will \vero{be trained to recognise} the pattern of dependence structures via this \vero{summary} 
dataset.  

Due to the influence of the air temperature at 2 meters above the surface on \vero{assessing  climate} changes and on all biotic processes, especially in extreme, \vero{we will apply our methods to 
this case study, data come from the European Center for Medium- Range Weather Forecasts  ECMWF. } The second case study  is the rainfall \vero{amount}   recorded in the east coast 
of Australia. 
 
\vero{The paper is organised as follows. } Section \ref{sec:2} \vero{is} devoted \vero{to} the theoretical tools used in paper.  An overview of \vero{ Convolutional Neural Networks} concepts 
\vero{is} exposed  in Section \ref{sec:3}. Section \ref{sec:4}  \vero{is} devoted to  configure the  architecture of \vero{the CNN for classification of dependence structures. } \verob{Section \ref{sec:5} shows the performance of our designed CNN on simulated data. Applications to 
environmental data}: Air temperature and Rainfall events \vero{are presented}  in Section \ref{sec:6}. Finally, \vero{a  discussion and main conclusions of this study are given in }
Section \ref{sec:7}.     
  
\section{Theoretical tools} \label{sec:2}
\vero{Let us give a survey on spatial extreme models ad tail dependence functions, see \cite{coles1999dependence} for more details. }
\subsection{Spatial extreme models}\label{sec:2.1}
\vero{Let $\{X'_i (s)\}_{s\in\S}$, $\S\in\R^d, d\geq 1$ be i.i.d replications of a stationary process.}  Let $a_n(s)>0$ and 
$b_n(s)$, $n\in\N$ be two sequences of continues functions. If 
\begin{equation}
\{\max_{\forall i\in n}\big(X'_i(s)-b_n(s)/a_n(s)\big)\}_{s\in\S}
 \overset{d}{\to} \{X(s)\}_{s\in\S}
\end{equation}
as $n\to\infty$, with \vero{non-degenerated marginals, then  $\{X(s)\}_{s\in\S}$ is a max-stable process. Its marginals  are Generalized Extreme Value (GEV)}. If for all $n\in\N$, 
$a_n(s)=1$ and $b_n(s)=0$, then  $\{X(s)\}_{s\in\S}$ \vero{is} called a simple max-stable \vero{process. It  has unite Fr\'echet marginal, which means:} $\P r\{X(s)\leq x\}=\exp(-1/x),x>0$, (see 
\cite{de2006spatial}). \vero{In \cite{de1984spectral}, it is  proved} that any simple max-stable process defined on \vero{a} compact set $\S\subset \R^d, d\geq1$ with continuous sample path 
\vero{admits a  spectral representation as follows}.

Let $\{\xi_i, i\geq1\}$ be  an i.i.d Poisson point process on $(0,\infty)$, with intensity $d\xi/\xi^2$ and let $\{W_i^+(s)\}_{i\geq1}$ be a i.i.d replicates of \vero{a} positive random filed 
$W:=\{W(s)\}_{s\in\S}$, such that $\mathbb{E}[W(s)]=1$.  Then  
\begin{equation}\label{def:max-stable}
X(s):=\max_{i\geq1}\{\xi_iW_i(s)^+\}, s\in\S, \S\in\R^d, d\geq 1
\end{equation}   
\vero{is a simple max-stable process.} The multivariate distribution function \vero{is given by}
\begin{equation}
\P r\{X(s_1)\leq x_1,\cdots,X(s_d)\leq x_d\}=\exp(-V_d(x_1,...,x_d)),
\end{equation} 
where \veroc{  $s_1,\cdots,s_d\subset \/ \S$ and $V$ is called the} exponent measure. \vero{It is homogenous of order $-1$ and} has  the expression: 
\begin{equation}
V_d(x_1,...,x_d)=\mathbb{E}\big[\max_{j=1,\cdots,d}\{W(s_j)/x_j\}\big],
\end{equation}
The extremal dependence  coefficient \vero{is}  given by $\Theta_d=V_d(1,\cdots,1)\in[1,d]$. \vero{It has been shown  \naf{by \cite{schlather2003dependence}} that for max-stable processes, either $\Theta_d=1$ 
which means that the process is asymptotically dependent (AD) or $\Theta_d=d$ which is the independent case. If $\Theta \neq 1$, the process is said to be asymptotically independent (AI). For 
max-stable processes, AI implies independence. } \cite{wadsworth2012dependence} \vero{introduced inverted max-stable processes which may be AI without being independent. Let $\{X(s)\}_{s\in\S}$ be a 
simple max-stable process, an inverted max-stable process $Y$ is defined as} 
\begin{equation}\label{invers}
Y(s)=-1/\log\{1-\exp(-1/X(s))\}, s\in\S\/.
\end{equation}
\vero{It has unit Fr\'echet marginal laws and its multivariate survivor function  is }
\begin{equation}
\P r\{Y(s_1)>y_1,\cdots,Y(s_d)>y_j\}=\exp(-V_d(y_1,\cdots    ,y_d))\/.
\end{equation}

\vero{In the definition of max-stable processes, }different models \vero{for} $W$ \veroc{lead to different} simple max-stable models, as well as  inverted max-stable models.  For instance, 
\vero{the Brown-Resnick model is constructed with } $W_i(s)=\{ \epsilon_i(s)-\gamma(s)\}_{i\geq1}$, where $\epsilon_i(s)$ \vero{are} i.i.d replicates  \vero{of a }  stationary Gaussian process  
with zero mean and \vero{variogram $\gamma(s)$ (see \cite{brown1977extreme} and \cite{kabluchko2009stationary}}. Many other models \vero{have been} introduced, such as Smith, Schlather and 
Extremal-t  introduced respectively by \cite{smith1990max}, \cite{schlather2002models} and \cite{opitz2013extremal}.  \\
\verob{In what follows, we shall consider extreme Gaussian processes which are Gaussian processes whose marginals have been turned to a unite Fréchet distribution. We shall also consider max-mixture processes which are $\max(aX(s)\/,(1-a)Y(s))$ where $a\in [0\/,1]$, $X(s)$ is a max-stable process and $Y(s)$ is an invertible max-stable process or an extreme Gaussian process.}

\subsection{Extremal dependence  measures}
Consider  a stationary spatial process $X:=\{X(s)\}_{s\in\S}$, $\S \subset \R^d, d\geq2$. \vero{The upper and lower tail dependence functions have been constructed in order to quantify the strength 
of AD and AI respectively.  The \veroc{upper tail} dependence coefficient $\chi$ is introduced in \cite{ledford1996statistics} and defined by
 \begin{equation}\label{ext:1}
\chi(h)=\lim_{u\to1}\P\big(F(X(s))>u|F(X(s+h))>u\big)\/, 
\end{equation}
where $F$ is the marginal distribution function of $X$. If $\chi(h)=0$, the pair $(X(s+h)\/,X(s))$ is  asymptotically independent (AI). If $\chi(h)\neq0$, the pair $(X(s+h)\/,X(s))$  is 
asymptotically dependent (AD). The process is AI (resp. AD) if  $\exists h \in\S$ such that $\chi(h)=0$  (resp. $\forall h\in\S \/, \  \chi(h)\neq 0$). } 
%
\vero{In \cite{coles1999dependence}, the lower tail dependence coefficient $\overline{\chi}(h)$  is proposed  \veroc{in order} to study the strength of dependence in AI cases. It is defined as:
\begin{equation}\label{ext:4}
\overline{\chi}(h)=\lim_{u\to1}\left[\frac{2\log\P\big(F(X(s))>u\big)}{\log\P\big(F(X(s))>u,F(X(s+h))>u\big)} -1\right] \/,\quad 0\leq u\leq1.
\end{equation} 
We have $-1\leq \overline{\chi}(h)\leq 1$ and the spatial process is AD if  $\exists h\in\S$ such that  $\overline{\chi}(h)=1$. Otherwise, it is a AI.}\\
\vero{Of course, working on data require to have empirical versions of these extreme dependence measures. We denote them respectively by $\hat{\chi}$ and $\hat{\overline{\chi}}$, \verob{they have been defined in \cite{wadsworth2012dependence}, see also \cite{bacro2016flexible}.}} \naf{Consider $X_i, i=1,2,\cdots, N$ the copies of \veroc{a spatial process $X$},  the corresponding empirical versions of $\chi(h)$ and $\bar{\chi}(h)$ \verob{are respectively} 
\begin{equation} \label{emp.chi}
\verob{\hat{\chi}(s\/,t)}=2-\frac{\log\big(N^{-1}\sum_{i=1}^N\mathbb{1}_{\{\hat{U}_i(s)<u,\hat{U}_i(t)<u\}}\big)}{\log\big(N^{-1}\sum_{i=1}^N\mathbb{1}_{\{\hat{U}_i(s)<u\}}\big)}, 
\end{equation}
and 
\begin{equation}\label{emp.chihat}
\verob{\hat{\bar{\chi}}(s\/,t)}=\frac{2\log\big(N^{-1}\sum_{i=1}^N\mathbb{1}_{\{\hat{U}_i(s)>u\}}\big)} {\log\big(N^{-1}\sum_{i=1}^N\mathbb{1}_{\{\hat{U}_i(s)>u,\hat{U}_i(t)>u\}}\big)}-1,
\end{equation}
where  $\hat{U}_i:=\hat{F}(x_i)=N^{-1}\sum_{i=1}^N\mathbb{1}_{\{X\leq x_i\}}$, \verob{for $|s-t|=h$}.}
 \section{Convolutional Neural Network (CNN) }\label{sec:3}
\verob{A}  Convolutional Neural Network CNN is an algorithm constructed and perfected to be one of the primary branches in deep learning. \vero{We shall use this method in order to recognise the dependence 
structure in spatial patterns}.  It stems from two studies introduced by \cite{hubel1968receptive} and \cite{fukushima1980neocognitron}. CNN \vero{are} used in many domains, one of the common 
use is \vero{for} image analysis. It \vero{appears to be} relevant in \vero{order to identify} the dependencies between nearby pixels (locations). It \vero{may recognise} spatial features 
\vero{(see \cite{wang2019deep})}.  Mainly, \vero{a} convolutional neural network consists \vero{in} three basic layers: Convolutional, pooling, and fully connected. The first two layers 
\vero{are} dedicated to feature learning and the latter one for classification. \vero{Many} researches \veroc{present the CNN architecture, see e.g.} \cite{yamashita2018convolutional} \verob{or   \cite{caterini2018deep}  for a mathematical framework }.  
%
\verob{We shall not provide the details of the CNN architecture, as it exists in many articles and books, we refer the interested reader to the above references. Let us just recall that, reconsidering spatial data, a convolution step is requiered. It is helpfull to make \veroc{the procedure} invariant by translation. \\
Once the CNN is build, the kernel values of the \veroc{convolutional layers} and the weights of the fully connected layers are learned during a training process. 

}
Training is the process of adjusting \vero{ values of the kernels and weights using known} categorical datasets.  The process has two steps, the first one is the forward propagation and the second 
step is the backpropagation. In forward propagation, the network performance  \veroc{is evaluated} by a loss function according to the kernels and weights updated in the previous step. From the 
value of the loss, the kernel and \veroc{weights is updated by a } gradient descent optimisation algorithm. If the difference between the true and predicted class of the dataset \vero{is} 
acceptable, \vero{the} process of training \veroc{stops}. \vero{So} that, selecting the suitable loss function and gradient descent optimisation algorithm is decisive \veroc{for} the quality of the constructed 
network performance.  Determining the loss function (objective function) should be \vero{done} according to the network task. \vero{Since our goal consists in classification, we shall use 
the cross-entropy as an objective function to minimize}. \naf{Let $y_a, a=1,\cdots A$ be the true class (label) of the dataset and \verob{let $\rho_a$ be the estimated probability of the $a$-th class, the cross-entropy loss function can formulated as $$L=-\sum_{a=1}^{A} y_a\log(\rho_a).$$ } }
Minimizing the loss means updating the  \veroc{parameters}, i.e. kernels and weights until \verob{the} CNN predicts the \vero{correct} class. This update will \vero{be} done by the gradient descent optimization algorithm in  \vero{the} \veroc{back propagation} step.  
Many gradient  algorithms 
\vero{are} proposed. The most commonly used in CNN are stochastic gradient descent (SGD) and Adam algorithm \cite{kingma2014adam}.  \verob{This algorithm is also an hyper-parameter in the network.} 
To begin \vero{ the training process, the data is } divided into three parts. The first part is devoted to training CNN. Monitoring the model performance, hyperparameter tuning and model selection 
\veroc{are done with}  the second part. This part \vero{is} called validation dataset.  \vero{This third part is used} for the evaluation of the final model performance. This latter part of data \vero{has} never \veroc{been} seen before \vero{by the}  CNN.

\section{Configure CNN to classify the types of dependence \vero{structures} }\label{sec:4}
\vero{We shall now explain how we used the CNN \verob{technology}  for our purpose to distinguish extreme dependence structures in spatial data.} 
\subsection{Constructing  the dependence  structures of the event}
\begin{figure}[h]
    \centering
       \includegraphics[scale= 0.75]{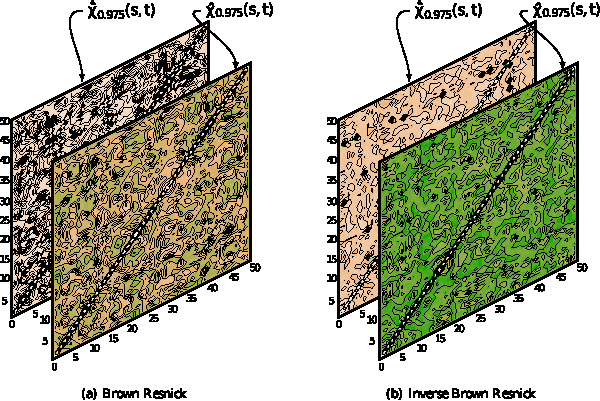} 
    \caption{Spatial dependence structures (layers) of  data samples generated from: (a) Brown-Resnick \verob{- asymptotic dependent model} ; (b)  inverted Brown-Resnick \verob{ - asymptotic independent model. The two models are generated with scale and smooth parameters  $0.4$ and $0.7$, receptively. Two tensors are  the empirical upper tail dependence measure $\hat{\chi}_{0.975}(s,t)$ and the empirical lower tail dependence measure $\hat{\bar{\chi}}_{0.975}(s,t), s,t=1,...,50)$, respectively}. }\label{input}
\end{figure}

\vero{Spatial extreme models may have} different dependence \vero{structures, such as asymptotic dependence or asymptotic independence}. These structures \vero{may be identified} by 
many measures. The well known measures able to capture  these structures are \vero{e.g.} upper and lower dependence measures  $\chi$ and $\bar{\chi}$ in \vero{Equations (\ref{ext:1}) and 
(\ref{ext:4})}, respectively. Upper tail  is \vero{able} to capture the dependence structure for asymptotic dependence models, but it fails  with  \vero{asymptotically independent models}.  The lower 
tail measure \veroc{treats}  this problem by \vero{providing the dependence strength for asymptotically independent models}.  \verob{We propose to consider these two measures $\chi$ and $\overline{\chi}$ as learning data for the CNN}, because each of  them provide information \vero{on} each type of dependence structure.  \vero{ The empirical counterparts  $\hat{\chi}(h)$ and $\hat{\bar{\chi}}{(h)}$ \naf{in Equations (\ref{emp.chi}) and (\ref{emp.chihat}) respectively}, computed above a threshold $u$ will \verob{be}  used } on the raw data to construct raster dataset with a symmetric array in two tensors, the first one for $\hat{\chi}$  
and the second for $\hat{\bar{\chi}}$.  This array will present the dependence structure of the corresponding data. Figure \ref{input} \vero{shows} of an array constructed   from Brown-Resnick and 
inverse Brown-Resnick models. 

\subsection{Building CNN architecture }
\begin{figure}[h]
    \centering
    \includegraphics[width=0.8\textwidth]{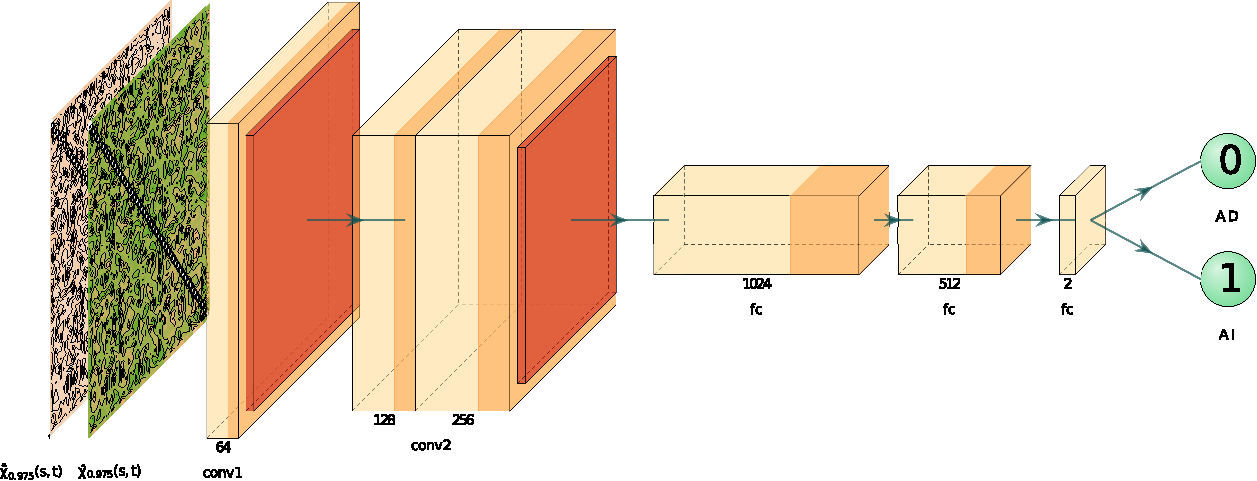}
    \caption{CNN architecture designed for classification asymptotic dependence  and asymptotic independence models. The \verob{input data is the} dependence structure array with two tensors, one for $\hat{\chi}_{0.975}(s,t)$ and the second for $\hat{\bar{\chi}}_{0.975}(s,t)$. Three convolutional layers with two max-pooling and fully connected  layers \verob{are the main parts of the CNN}. The third layer \verob{is} devoted to classification.  \label{CNN:art}}
\end{figure}

It is essential  \verob{in  the convolutional neural network design to take into account  the kind of data and the task to be done}: classification, representations, \verob{or} anomaly detection.  
Practically, designing \verob{a CNN for classification of complex patterns remains challenging. \veroc{First of all}, one has to determine the number of convolutional and fully connected layers that shoud be used. Secondly, tuning a high number of parameters (kernel, weights) is required.  Many articles are devoted to build and improve CNN architectures to have good performance: \cite{lecun1990handwritten},  \cite{NIPS2012_4824},  \cite{he2016deep} and \cite{xie2017aggregated}}.\\
\verob{These} CNN architectures \verob{are  appropriate for image classifications but not for our goal because we need to keep the dependence structure. So that, we designed a CNN for our dependence }classification aim.  \veroc{From many attempts, we found out} that \verob{out that quite a high number of  parameters is required: not} less than $17$ million parameters.  Figure \ref{CNN:art} shows the general framework  of the  CNN architecture  designed for the \verob{dependence} classification. 

The Input of this designed CNN \veroc{is} the dependence structure layers, consisting \verob{in} two tensors, \verob{one represents $\hat{\chi}(s,t)$ and the other one represents} $\hat{\bar{\chi}}(s,t)$, where  $s,t=1,\cdots 50$. 

Two networks \verob{are constructed}, one has two classes output \veroc{called} 2-class, for recognizing  asymptotic dependence vs asymptotic independence  dependence structure. \verob{The second CNN has  a third ouput class in order to detect if a spatial process is neither AD nor AI. The third class is considered  as unknown dependence structure type}.  Table \ref{tab:2}, shows \verob{the} details of the  architectures. 

\begin{table}[h]
\caption{Designed Convolution Neural Network architecture \verob{for the two classes output. For the CNN with three output classes, the architecture is the same but the last fully connected layer has} three units rather than two.}\label{tab:2}
\centering
     \small\addtolength{\tabcolsep}{-3pt}
     \normalsize
\begin{tabular}{lccclc}
\hline
\hline
\textbf{Layer type} & \textbf{Feature Map} & \textbf{Size of Kernels} & \textbf{Stride size} &  \textbf{Padding} & \textbf{Activation} \\
\hline
\hline
Input               & --                   & --                       & --                   & --      &                 --                  \\
2D-Convolutional    & 64                   & $3\times 3$              & $2\times 2$          & Valid   & ReLU                \\
2D-Max Pooling      & --                   & $2\times 2$              & $1\times 1$          & Valid   & --                  \\
2D-Convolutional    & 128                 & $3\times 3$              & $1\times 1$          & Valid   & ReLU                \\
2D-Convolutional    & 256                  & $3\times 3$              & $1\times 1$          & Valid   & ReLU                \\
2D-Max Pooling      & --                   & $2\times 2$              & $1\times 1$          & Valid   & --                  \\
Fully Connected     & 1024                 & --                       & --                   &         & ReLU                \\
Fully Connected     & 512                 & --                       & --                   &         & ReLU                \\
Output      & 2                 & --                       & --                   &         & Softmax             \\

\hline
\hline
\end{tabular}
\end{table}

\verob{A regularizer with regularization factor  $l_2=0.00005$ is} added to each convolutional layer. The gradient rate \verob{is set to be $\alpha=0.0001$} when updating the weights of the model. The total parameters for this architecture \veroc{is more than $17$ million and $45$ million for datasets with $30$ and $40$ locations, respectively}. In deep learning, \verob{the choice of an}  optimization algorithm is crucial \verob{in order to reach} good results. Adam optimization algorithm is very effective with CNN (\verob{see  \cite{kingma2014adam}}). In this study, Adam optimization algorithm with a learning rate $\lambda=0.0001$ has been used.  Since the dataset  are categorical,  \verob{the cross-entropy objective} function is more suitable.  \verob{\tt{Keras} package in R interface} \veroc{is used for model learning}. 

\section{Evaluation of the performance of CNN via  simulation }\label{sec:5}
\verob{In order to} evaluate the performance of the constructed CNN networks in the previous section, three scenarios have \verob{been} applied. For each scenario, \verob{the} $2$-class and $3$-class networks \verob{are trained on AD and AI processes, for the $3$ class networks, max-mixture processes \veroc{(see definition in Section \ref{sec:2.1})} are added to the training data. Our training data consists in: 
\begin{itemize}
 \item max-stable processes \veroc{(defined in Equation (\ref{def:max-stable}))} with $1\/000$ observations on sites  $s_i, \ i=1,\cdots 30$,  $60\/000$ datasets are  generated from four spatial extreme models: Smith, Schather, Brown-Resnick and Extremal-t with  scale and smooth parameters $\sigma$ and $\delta$ respectively.  These parameters are either chosen at random or in regular sequences,
 \item inverse max-stable as defined in (\ref{invers}) with the same parameters as above and  $5\/000$ datasets are generated from extreme Gaussian processes in order  to have more variety in AI models. 
\end{itemize}
}
In total, for the two dependence structure  types $125\/000$ datasets \verob{are generated and divided into} three parts, $64\%$ for training, $16\%$ for validation and $20\%$ for testing.  Empirical \verob{$\hat{\chi}_{0.975}(s_i,t_i)$ and $\hat{\bar{\chi}}_{0.975}(s_i,t_i)$ with $ (s_i,t_i)\in[0,1]\/, i=1,\cdots 30$ defined in (\ref{emp.chi}) and (\ref{emp.chihat}) respectively are used to summarize the datasets and are the inputs for training the CNN. For the  $3$-class network, we  added $12\/000$ datasets  with neither AD nor AI dependence structure, through max-mixture processes. We have performed several scenarios.} 
\begin{itemize}
\item In the first scenario, for each dataset,  the locations $s\in[0,1]^2 $ are \verob{uniformly randomly chosen}. Moreover  the \verob{scale and smoothness} parameters are also \verob{uniformly randomly selected: $\sigma\sim U(0,1)$ and $\delta\sim U(1,1.9)$}. In the $3$-class network, the mixing parameter \verob{$a$ is also uniformly randomly selected:  $a\sim U(0,1)$. The AD and AI models in the max-mixture are also chosen at random in the different classes. }
\item \verob{In the second scenario}, the locations was fixed for all datasets, and the parameters \verob{remain chosen at random.}
\item  In the third \verob{scenario}, the locations are fixed for all datasets and the parameters \verob{ran through regular sequences, $\sigma\in [{0.1}\/,{1}]$ and $\delta\in [0.1\/, 1.9]$, with steps $0.2$, the mixing parameter $a\in [0.3\/,0.7]$ with steps $0.1$.} 
\end{itemize}
\verob{The evaluation task is done for the three scenarios described above. The datasets used are AD or AI for the $2$-class networks and we added max-mixture processes for the $3$-class networks. For the random scenarios, the evaluation datasets sites and parameters are chosen at random. For the fix locations scenario, evaluation datasets sites are chosen different from the training ones. For scenario $3$, scale and smoothness parameters are chosen different for evaluation and training. } The losses and accuracy for all datasets considered for the evaluation \verob{are} shown  in Table \ref{tab:simu}.  
\begin{landscape}
\begin{table}[]
\caption{The losses and Accuracy of 2-class and 3-class CNN networks trained \verob{within} the three scenarios.  }\label{tab:simu}
\centering
     \small\addtolength{\tabcolsep}{-3pt}
     \normalsize
\begin{tabular}{|c|c|c|c|c|c|c|c|c|c|c|c|c|}
\hline
                                                                       & \multicolumn{4}{c|}{\begin{tabular}[c]{@{}c@{}}Scenario 1\\ Fixed locations \\ and \\ random parameters\end{tabular}}                                           & \multicolumn{4}{c|}{\begin{tabular}[c]{@{}c@{}}Scenario 2\\ Random locations \\ and \\ random parameters\end{tabular}}                                          & \multicolumn{4}{c|}{\begin{tabular}[c]{@{}c@{}}Scenario 3\\ Fixed locations \\ and \\ sequential parameters\end{tabular}}                                       \\ \hline
                                                                       & \multicolumn{2}{c|}{\begin{tabular}[c]{@{}c@{}}2-class\\ Network\end{tabular}} & \multicolumn{2}{c|}{\begin{tabular}[c]{@{}c@{}}3-class\\ Network\end{tabular}} & \multicolumn{2}{c|}{\begin{tabular}[c]{@{}c@{}}2-class\\ Network\end{tabular}} & \multicolumn{2}{c|}{\begin{tabular}[c]{@{}c@{}}3-class\\ Network\end{tabular}} & \multicolumn{2}{c|}{\begin{tabular}[c]{@{}c@{}}2-class\\ Network\end{tabular}} & \multicolumn{2}{c|}{\begin{tabular}[c]{@{}c@{}}3-class\\ Network\end{tabular}} \\ \hline
                                                                       & Loss                                  & Accuracy                               & Loss                                  & Accuracy                               & Loss                                  & Accuracy                               & Loss                                  & Accuracy                               & Loss                                  & Accuracy                               & Loss                                  & Accuracy                               \\ \hline
Training                                                               & 0.3255                                & 0.8668                                 & 0.6829                                & 0.7232                                 & 0.3649                                & 0.8386                                 & 0.3740                                & 0.8749                                 & 0.2985                                & 0.8824                                 & 0.4729                                & 0.8242                                 \\ \hline
validation                                                             & 0.3525                                & 0.8564                                 & 0.7071                                & 0.7108                                 & 0.3675                                & 0.8406                                 & 0.4763                                & 0.8411                                 & 0.3285                                & 0.8658                                 & 0.5204                                & 0.8018                                 \\ \hline
testing                                                                & 0.3582                                & 0.8547                                 & 0.7080                                & 0.7146                                 & 0.3694                                & 0.8400                                 & 0.5165                                & 0.8039                                 & 0.3275                                & 0.8672                                 & 0.5169                                & 0.8041                                 \\ \hline
Gaussian                                                               & 0.0700                                & 0.9969                                 & 0.0928                                & 1.0000                                 & 0.2851                                & 0.9550                                 & 0.3868                                & 0.9540                                 & 0.0658                                & 0.9980                                 & 0.0655                                & 0.9980                                 \\ \hline
Asymptotic dependent                                                   & 0.4085                                & 0.7880                                 & 0.7052                                & 0.6870                                 & 0.4796                                & 0.7050                                 & 0.6378                                & 0.6660                                 & 0.3772                                & 0.7940                                 & 0.5493                                & 0.7480                                 \\ \hline
Asymptotic independent                                                 & 0.2889                                & 0.9390                                 & 0.5124                                & 0.8970                                 & 0.2753                                & 0.9610                                 & 0.4621                                & 0.9120                                 & 0.3284                                & 0.9020                                 & 0.3317                                & 0.9240                                 \\ \hline
\verob{Mixtures}                                                                & ---                                   & ---                                    & 0.7437                                & 0.6510                                 & ---                                   & --                                     & 6.0335                                & 0.1370                                 & ---                                   & ---                                    & 0.8299                                & 0.6633                                 \\ \hline
Different locations                                                    & 0.779                                 & 0.8060                                 & 0.8888                                & 0.8000                                 & ---                                   & ---                                    & ---                                   & ---                                    & 0.9536                                & 0.8010                                 & 1.1231                                & 0.7990                                 \\ \hline
\begin{tabular}[c]{@{}c@{}}Different  scale \\ parameters\end{tabular} & ---                                   & ---                                    & ---                                   & ---                                    & ---                                   & ---                                    & ---                                   & ---                                    & 0.3908                                & 0.8480                                 & 0.5024                                & 0.8100                                 \\ \hline
\begin{tabular}[c]{@{}c@{}}Different smooth \\ parameters\end{tabular} & ---                                   & ---                                    & ---                                   & ---                                    & ---                                   & ---                                    & ---                                   & ---                                    & 0.3295                                & 0.8744                                 & 0.4361                                & 0.8355                                 \\ \hline
\end{tabular}
\end{table}

\end{landscape}

\begin{figure}[h]
    \centering
       \includegraphics[scale= 0.5]{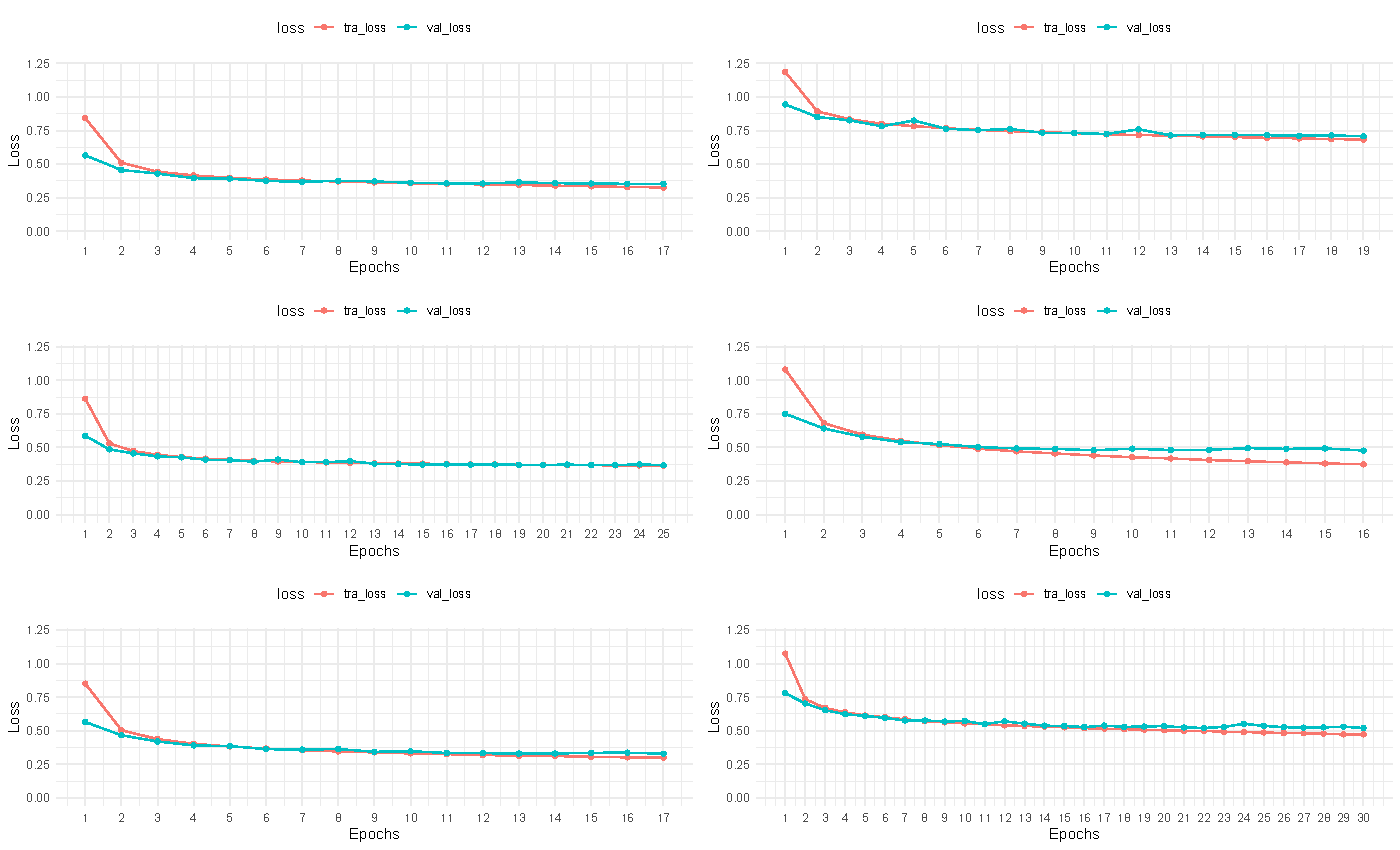} 
    \caption{\verob{Loss for training and validation} for each scenario and class network. \verob{Each row represents the progress for} scenarios  $1$,$2$ and $3$, respectively. While the columns \verob{represent the process} of the $2$-class and $3$-class networks, respectively. }\label{prog_6}
\end{figure}

\verob{The training progress is illustrated  in} Figure \ref{prog_6}.

Regarding the general performances of the scenarios illustrated in Figure \ref{prog_6}, the training progress of all networks \verob{is correct}, the training and validation losses \verob{decrease, we observe no under nor overfitting and, moreover the procedure is stable. The network training will stop when there is no more improvement on the} validation loss. 

\verob{The first three rows of Table \ref{tab:simu} show the training validation and testing losses. We can conclude that both $2$-class and $3$-class networks perform well for the $3$ scenarios. The generalization is better with the third scenario.  The performance of the networks may also be examined specifically for different dependence structures.}
Asymptotic independence structure (inverse max-stable \verob{or extreme Gaussian}) \verob{is recognized almost perfectly by all} networks as shown in the Table \ref{tab:simu}. \verob{The performance in recognizing the asymptotic dependent structures is less} satisfactory.  \verob{The best results are obtained for Scenario 3, both for  $2$ and $3$-class networks. The mixed dependence structure may be recognized by the 3-class networks:  the second scenario failed to distinguish it, while the two other scenarios provide acceptable results.  For the different location tests, the networks trained with  scenario $1$ overcome those trained with scenario $3$. This is because \veroc{the networks} trained with random $\sigma$ and $\delta$ parameters lead to different dependence structures. In other words, no specific dependence structures are learned, contrary to the third scenario. The performances are much improved by training the networks with  datasets whose parameters cover sequences of parameters. Finally, scenario  $3$ has good performances, even  for datasets with untrained scale and smooth parameters. These observations lead us to use the third scenario in our application studies: $2$-meters air temperature over Iraq and the rainfall over the East coast of Australia.  }

\section {Application to the environmental case studies}\label{sec:6}
\verob{Modeling spatial extreme dependence structure on environmental data is our initial purpose in this work.  We finish this paper wtih two specific studies on Iraqi air temperature and East Austrian rainfall. }
  \subsection{Spatial dependence pattern of the air temperature at two meter above Iraq land}
\verob{Temperature of the air at two meters above the surface has a major influence on assessing climate changes as well as on all biotic processes.  This data is inherently a spatio-temporal process (see \cite{hooker2018global}). }
  \subsubsection{The data}
\begin{figure}[!h]
    \centering
         \includegraphics[width=\textwidth]{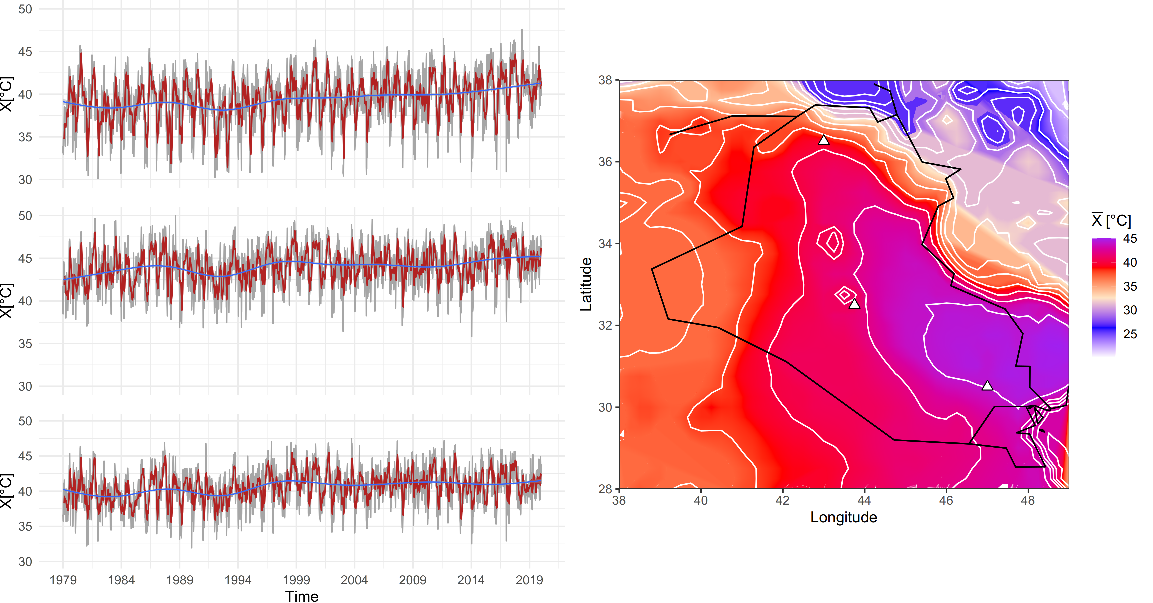} 
    \caption{Left panel, (gray lines) represent the time series of the daily  average of  the 2 meter air temperature for the period 1979-2019 verob{along summer months  (June, July and August). The red lines} represent  the simple 10 days moving average. The smoothing temporal data is in blue line. The contour plot in the \verob{right} panel shows the gradient level in the mean of $X$ for the entire period above the Iraq land.}  
\label{TS}
\end{figure}

\verob{We used data}  produced by  the meteorological reanalysis ERA5 \verob{and} achieved by the European Center for Medium-Range Weather Forecasts  ECMWF. An overview  and quality assessment of this data \verob{may be found in}  \url {http://dx.doi.org/10.24381/cds.adbb2d47}. \verob{Our} objective  is to \verob{study}  the spatial dependence structure pattern of this data  \verob{recorded from a high temperature region in Iraq. }
Let $\{X_k(s)\}_{s\in\S, k\in \K}$, $\S\subset \R^2,k\in \K\subset \R^+$, be the daily average of the 2 meter air temperature process computed  at the peak hours from 11H to 17H for the period 1979-2019 along of the summer (June, July and August ). This collection of data  \verob{results in $|\K|=3772$  temporal replications and $|\S|=1845$} grid cells. \verob{The data has naturally a spatio-temporal nature. \veroc{Nevertheless, a preliminary preprocessing suggests} us to treat them as independent replications of a stationary spatial process. }
The left panel in Figure \ref{TS} shows the  time series of the  $X$  for three locations located in the north, middle and south of Iraq (white triangles on the right panel). The right panel, shows the \verob{temporal mean}.   

Regarding the time series in the left panel, the data in three locations \verob{may be considered as stationary in time}.  
\verob{In order to remove the spatial non stationarity, we shall aply a simple moving average, as used in \cite{huser2020eva}, see Section  \ref{anomaly:1}.}

\subsubsection{\verob{Preprocessing of $2$ meter air temperature data.}}\label{anomaly:1}
\begin{figure}[!h]
    \centering
         \includegraphics[width=\textwidth]{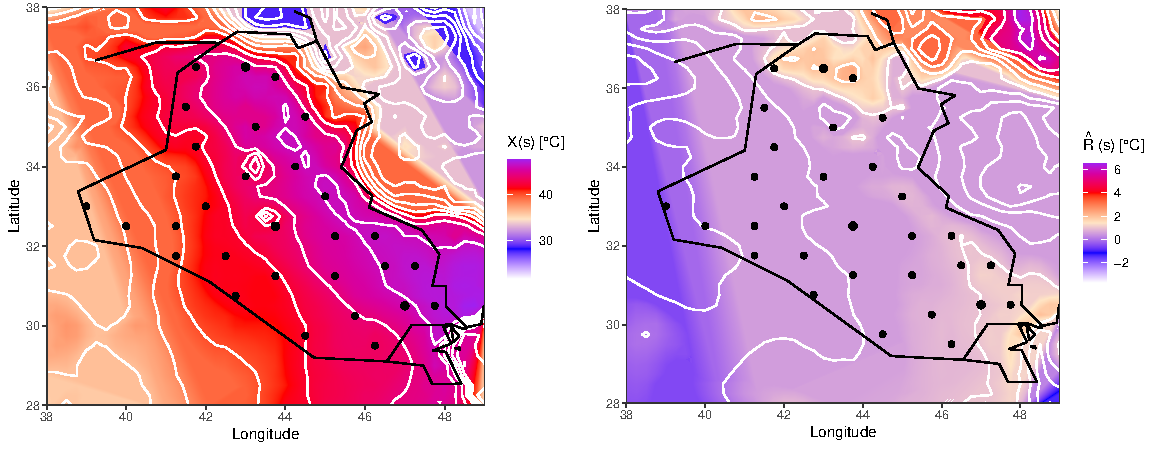}  
    \caption{Two meter air temperature $X_k(s)$ over Iraq at \verob{August 15th, 2019 in the left panel, while the estimated residual  process $\hat{R}_k(s)$ is } in the  right panel. \veroc{The black dots are the locations} chosen to construct the air temperature dependence structure.}  
\label{smoothness}
\end{figure}

\verob{As mentioned above, the $2$ meter summer air temperature data in Iraq look spatially non stationary. We propose to follow \cite{huser2020eva}, in order to remove the non stationarity.
We shall decompose the spatial process $\{X(s)\}_{s\in\S}$ into two terms, the average part $\mu(s)$ and the residual part $R(s)$, so that  
\begin{equation}\label{eq:residual}
  X(s)= \mu(s)+R(s).
\end{equation}
}
Smoothing the empirical estimation of \verob{${\mu}_k(s)$ by a moving average over $10$ days leads to}
$$ \hat{R}_k(s)= X_k(s)-\hat{\mu}_k(s), s\in\S, k\in\K.$$ 
Figure \ref{smoothness},  \verob{shows the spatial variability for August, 15th 2019. One sees the non stationarity of $(X_k(s))_{s\in \S}$, while the residuals $\hat{R}_k(s)$ seem stationary (right panel)}. 

\begin{figure}[!h]
  \centering
         \includegraphics[width=0.75\textwidth]{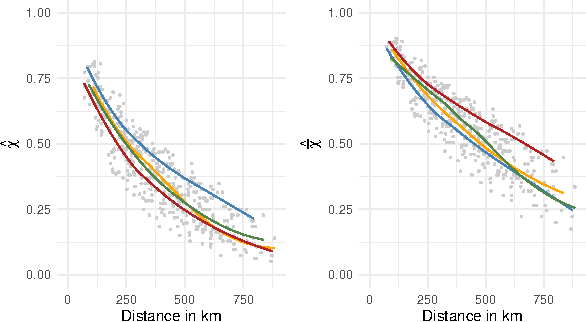}  
    \caption{\verob{Empirical tail dependence measures $\hat{\chi}_{0.975}(h)$ and $\hat{\bar{\chi}}_{0.975}(h)$, for each  direction. The }red line is for direction  $(-\pi/8,\pi/8]$, blue for $(\pi/8,3\pi/8]$, green for $(3\pi/,5\pi/8]$ and black for $(5\pi/8,7\pi/8]$, where $h=\|s-t\|, s,t\in\S$.  The gray dots represent the pairwise $\hat{\chi}_{0.975}(s,t)$ and $\hat{\bar{\chi}}_{0.975}(s,t)$ for all dataset.}  
\label{fig:isotropy}
\end{figure}

\verob{In model (\ref{eq:residual}), the residual process carries the dependence structure, we study below the isotropy of it. Figure \ref{fig:isotropy} shows the estimated tail dependence functions with respect to some directions (where $0$ is the north direction). From this graphical study, we may retain the istropy hypothesis.  }

\verob{We shall estimated the extremal dependence structure on block maxima.
Let $m\in \K$, $k\in \S$ and let }
$$\mathcal{B}_{m,k}(s)= \{(s,k):(k-m)\leq k\leq (k+m)\}\cap(\S\times\K)$$ 
be a temporal neighborhood set of  $\hat{R}_k(s)$ for each grid cell $s$, the extreme spatial process \verob{is} defined as 
$$Y_k(s)=\max_{(s,k^*)\in\mathcal{B}_{m,k}(s)}\hat{R}_{k^*}(s).$$ 
\verob{Then,  dependence structure of the air temperature will be estimated using $\hat{\chi}_{0.975}(s,t)$ and $\hat{\bar{\chi}}_{0.975}(s,t)$, $s,t=1,...,30, (s,t)\in\S$. We shall consider  the rank transformation applied  on $Y$, in order to transform the margin to unite Fr\'echet:}~ 
\[ \hat{Y}_k(s) =
  \begin{cases}
       -1/\log(Rank(Y_k(s))/(|\mathcal{B}_{m,k}(s)|+1) \big)      & \quad \text{if} Rank(Y_k(s))>0\\
 0 & \quad \text{if } Rank(Y_k(s))=0,
  \end{cases}
\]
where $k=1,...,|\mathcal{B}_{m,k}(s)|$, $s=1,...,30$ and $|\cdot |$ is the cardinality, \verob{we get a $[30\times 30\times 2]$ table which will consists in the CNN inputs. }
\subsubsection{Training the designed Convolutional Neural Network}
\begin{figure}[!h]
    \centering
        \includegraphics[scale=0.75]{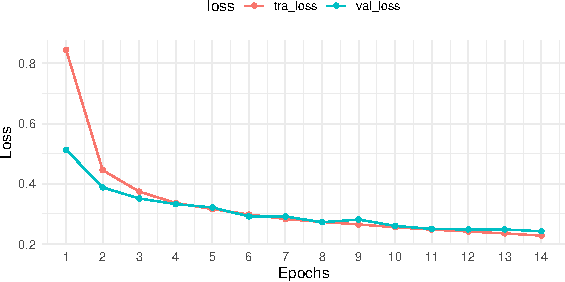}     
    \caption{The loss of training and validation recorded for each of the 14 epochs during the training progress. }  
\label{loss}
\end{figure}

\verob{We shall now use the CNN procedure described in Sections \ref{sec:4} and \ref{sec:5}, we consider the locations of the data,  according  to scenario $3$. We resize the data into $[0,1]^2$. The training datasets are generated  as in scenario $3$,  for parameters, we use regular sequences with steps $0.1$}. Figure \ref{loss}, shows \verob{the loss of the training progress for the designed CNN  $2$-class network. The performance of the $3$-class network is comparable. } 

As we mentioned previously,  the CNN will stop on training when the validation loss reaches the minimum. At the epoch $14$,  the training and validation loss recorded at $0.2282$ and $0.2427$ respectively, \verob{with accuray}  $0.9333$ and $0.9321$ respectively.  For teasing data the loss was $0.2512$ and accuracy $0.9260$. \verob{This shows that the training process worked well.}   \\

\subsubsection{\verob{Predicting the dependence structure class for two meter summer air temperature }}
The prediction of the pattern of \verob{ air temperature dependence structure is } done according to  7 sizes of block maxima, $m=192\/,\ 30\/,\ 15\/,\ 7\/,\ 5\/,\ 3\/,\ 1$ \verob{which gives} respectively  $41,125,251,538,754,1257$ blocks from \verob{$K=3772$ measurements, so that we shall see the influence of block} size on the predicted class. Table \ref{tab:3} \verob{shows} the predicted  pattern  of the dependence structure of 2 meter air temperature  corresponding to each block maxima size proposed. For all  \verob{block} sizes, the predicted pattern  was asymptotic dependence, \verob{with no significant effect of block size on} the probability of prediction \verob{for both $2$-class and $3$-class CNN. So, we may conclude that  the $2$ meter air summer temperature has an asymptotic dependent spatial structure}.
\begin{table}[!h]
\centering
   \small\addtolength{\tabcolsep}{-3pt}
     \caption{\verob{Predicted class and its probability for} the \veroc{$2$ meter air temperature data for each block maxima} size proposed. The dependence structure \verob{is validated by the two CNN. AD and AI refer}  to asymptotic dependence and asymptotic independence, respectively. }\label{tab:3}
          \normalsize
\begin{tabular}{lccccc}
\hline
\hline   
                   & \multicolumn{2}{c}{2 classes CNN}                                               & \multicolumn{3}{c}{3 Classes CNN}                                              \\ 
Block Maxima size & \multicolumn{1}{l}{Probability of AD} & \multicolumn{1}{l}{Probability of AI} & \multicolumn{1}{l}{Probability  of AD} & \multicolumn{1}{l}{Probability  of AI} & \multicolumn{1}{l}{Probability  of mix }\\
\hline
\hline
$m=92$ days      & 1.000          & 0.000     & 1.000    & 0.000      &   0.000  \\
$m=30$ days      & 1.000          & 0.000     & 1.000    & 0.000      &   0.000  \\
$m=15$ days      & 0.990          & 0.001     & 0.645    & 0.355      &   0.000   \\
$m=7$ days        & 0.860          &0.140      & 0.791    & 0.199      &   0.010  \\
$m=5$ days        & 0.929          & 0.071     & 0.686    & 0.013      &   0.301  \\
$m=3$ days        & 0.864          & 0.136     & 0.702   & 0.085       &  0.213   \\
$m=1$ day          & 0.995          & 0.005     & 0.950   & 0.037       & 0.013   \\ 
\hline
\hline
\end{tabular}
\end{table}

\subsection{Rainfall dataset: case study in Australia}
\begin{figure}[!h]
    \centering
       \includegraphics[scale=1]{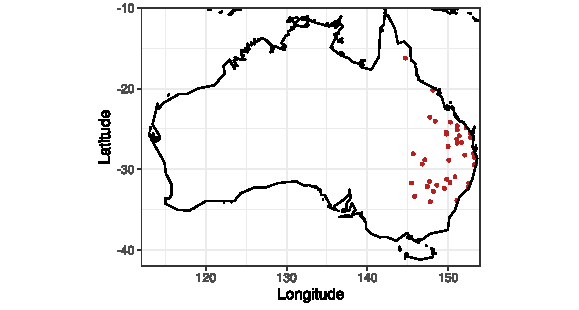}     
    \caption{Australia map illustrate the coordinates of rainfall  monitoring stations. }  
\label{austalia}
\end{figure}

\verob{Another dependence structure investigated in this paper is the daily rainfall  data recorded in 40 monitoring stations located in East of Australia illustrated by  the red dots in Figure \ref{austalia}.  }
\verob{This dataset has been studied by several authors \cite{bacro2016flexible,ahmed2017semi,abu2019fitting}. }
\subsubsection{The data}
For each location, the \verob{cumulative rainfall amount (in millimeters) over $24$ hours is recorded, during the period}  1972-2019 along the extended rainfall season (April-September). \verob{It results  $|\K|=8784$} observations. The locations \verob{have been selected} among many monitoring locations \verob{(red points on Figure \ref{austalia}),} keeping the elevations above mean sea level between $2$ to $540$ meters, \verob{in order to ensure} the spatial stationarity.  The data \verob{is} available freely on the website of Australian meteorology Bureau \url{http://www.bom.gov.au}. 
\verob{The spatial stationarity and isotropy properties have been investigated for the this data} in many papers, for instance see e.g., \cite{bacro2016flexible}, \cite{ahmed2017semi} and \cite{abu2019fitting}. \verob{We shall consider that the data is stationary and isotropic. }  
That leads to construct the corresponding dependence structure directly form the  data  itself without having to estimate the \verob{residuals as in the previous section}. Let $\{X_k(s)\}_{s\in\S,k\in\K}$, $s=1,...,40,k=1,...,8784$ be \verob{the}  spatial process \verob{representing the rainfall in the East cost} of Australia. Adopting the block maxima size \verob{as in the}  previous section, we \verob{consider} the extreme process:
$$Y_k(s)=\max_{(s,k^*)\in\mathcal{B}_{m,k}(s)}X_{k^*}(s)$$
\verob{and  transform $Y$ into a unite Fr\'echet marginals process. The} dependence structure of this \verob{data will be summarized in a $40\times 40\times 2$ array. The first and second} tensor are  $\hat{\chi}_{0.975}(s,t)$ and $\hat{\bar{\chi}}_{0.975}(s,t)$, $s,t=1,...,40$, respectively with  threshold $u=0.975$.

\subsubsection{\verob{Predicting the pattern of the dependence structure of rainfall amount in East Austria.} }
\begin{figure}[!h]
    \centering
        \includegraphics[scale=0.75]{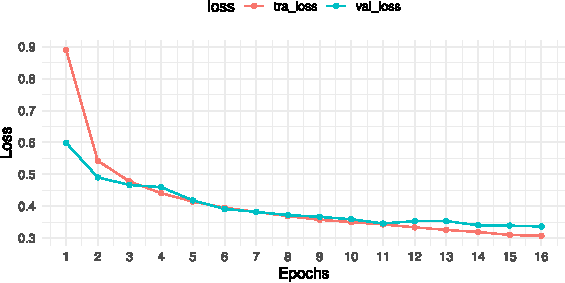}     
    \caption{\verob{Loss} of training and validation recorded for each of the 16 epochs during the training progress. }  
\label{loss_aus}
\end{figure}

\verob{We shall use} same designed CNN in the previous \verob{section}.
The training and validation progress \verob{are shown} in Figure \ref{loss_aus},

 The validation loss reached the minimum at epoch $16$, so the final performance of CNN will be \verob{calculated} at this epoch. The  training and validation  recorded  loss \verob{are} $0.307$ and $0.336$, respectively. The accuracy  \verob{are} $0.889$ and $0.875$. The loss and accuracy of \verob{the} tested data  were $0.342$ of loss and $0.870$ accuracy. Table \ref{tab:4} \verob{shows} the predicted class for each \verob{proposed block maxima size. }
 
 \begin{table}[!h]
\centering
     \small\addtolength{\tabcolsep}{-3pt}
     \caption{\verob{Predicted} class of the Rainfall data for each block maxima size proposed. The dependence structure \verob{is classified with} the two trained CNN. AD and AI refers  to asymptotic dependence and asymptotic independence, respectively.}\label{tab:4}
          \normalsize
\begin{tabular}{lccccc}
\hline
\hline   
                   & \multicolumn{2}{c}{2 classes CNN}                                               & \multicolumn{3}{c}{3 Classes CNN}                                              \\ 
Block Maxima size & \multicolumn{1}{l}{Probability of AD} & \multicolumn{1}{l}{Probability of AI} & \multicolumn{1}{l}{Probability  of AD} & \multicolumn{1}{l}{Probability  of AI} & \multicolumn{1}{l}{Probability  of mix }\\
\hline
\hline
$m=183$ days      & 0.020          & 0.980     & 0.020    & 0.980      &   0.000  \\
$m=30$ days      & 0.020           & 0.980     & 0.020    & 0.980      &   0.000  \\
$m=15$ days      & 0.060          & 0.940     & 0.063    & 0.937      &   0.000   \\
$m=10$ days      & 0.271          &0.729      & 0.143    & 0.808      &   0.049  \\
$m=5$ days        & 0.580          & 0.420     & 0.411    & 0.368      &   0.221 \\
$m=3$ days        & 0.746          & 0.254     & 0.000   & 0.009       &  0.991   \\
$m=1$ day          & 0.946          & 0.054     &0.000   & 0.001       & 0.999   \\ 
\hline
\hline
\end{tabular}
\end{table}

The classification \verob{procedure shows that the asymptotic independence structure is more suitable for block maxima sizes up to 10 days. This is in accordance with  \cite{bacro2016flexible} where rainfall amount} the same region, but for different locations \verob{was studied}. They concluded that it is not suitable to choose asymptotic dependence models for modeling seasonal maxima. 
While for block maxima \verob{of} size $m=5$ the prediction \verob{is} not decisive for \verob{the $2$-class CNN}. 
For 3 days and daily block maxima, the classifier with 2 classes gives a high probability for asymptotic dependence model. While,   CNN with 3 classes, gives different predictions:\verob{ with high probability  a mixture between AD and AI should be chosen}. 
Furthermore, the investigation for the same \verob{data has been done in} previous works \veroc{using different block maxima sizes, see \cite{bacro2016flexible}, \cite{ahmed2017semi} and   \cite{abu2019fitting}.  They  founded that max-mixture models are suitable. This is confirmed by the prediction of CNN with 3 classes.}

    \section {Discussion and conclusions }\label{sec:7}
Since the \verob{kind of dependence structures may have} influence on the nature of \verob{joined extreme events}, it is important to devote studies to this matter. Most of the studies \veroc{deal with  modeling }  extreme events directly by \verob{parametrical statistics methods, usually, without preliminary investigation on which pattern of dependence structure would be the most suitable}. Moreover, \verob{ the block maxima size has influence on the dependence structure.} 
In this paper,  far from the classical methods, \verob{we proposed to exploit the powerfulness  of Convolutional Neural Network}  to investigate the pattern of the dependence structure of the extreme events. Two \verob{environmental data (air temperature at two meters over Iraq and rainfall over the  East coast of Australia) have been studied in order to classify their dependence structures patterns.} The input of the designed CNN  \verob{are} the empirical upper and lower tail dependence measures  $\hat{\chi}_{0.975}(s,t)$ and $\hat{\bar{\chi}}_{0.975}(s,t)$. 
The training process \verob{has been done on} generated  data from  max-stable stable models,  \verob{inverse max-stable processes and extreme Gaussian processes, in order} to get asymptotic independent models. The data  \verob{are generated  according to fixed coordinates rescaled} in $[0,1]^2$.  
The ability of this model \verob{to recognize the pattern of dependence structure has been emphasized in  training, validation,  testing loss and accuracy. }

It is worth mentioning that the sensitivity of the dependence structure class by considering the size of block maxima should be taking into account \verob{in the models}. Adopting this classification procedure \verob{may} advise for choosing a reasonable size of block maxima such that the \verob{data has a} good representation. For instance, for the air temperature event, whatever the block maxima size is chosen, the dependence structure is asymptotic independence. 
While, for rainfall \verob{data}, the dependence structure class changed across \verob{block size}.




\section*{Acknowledgments}
This work was supported by PAUSE operated by Collège de France, and the LABEX MILYON (ANR-10-LABX-0070) of Université de Lyon, within the program ``Investissements d'Avenir'' (ANR-11-IDEX-0007) operated by 
the French National Research Agency (ANR). 










\bibliography{wileyNJD-APA}%

\begin{thebibliography}{}

\bibitem [\protect \citeauthoryear {%
Abu-Awwad%
, Maume-Deschamps%
\BCBL {}\ \BBA {} Ribereau%
}{%
Abu-Awwad%
\ \protect \BOthers {.}}{%
{\protect \APACyear {2019}}%
}]{%
abu2019fitting}
\APACinsertmetastar {%
abu2019fitting}%
\begin{APACrefauthors}%
Abu-Awwad, A.%
, Maume-Deschamps, V.%
\BCBL {}\ \BBA {} Ribereau, P.%
\end{APACrefauthors}%
\unskip\
\newblock
\APACrefYearMonthDay{2019}{}{}.
\newblock
{\BBOQ}\APACrefatitle {Fitting spatial max-mixture processes with unknown
  extremal dependence class: an exploratory analysis tool} {Fitting spatial
  max-mixture processes with unknown extremal dependence class: an exploratory
  analysis tool}.{\BBCQ}
\newblock
\APACjournalVolNumPages{Test}{}{}{1--44}.
\PrintBackRefs{\CurrentBib}

\bibitem [\protect \citeauthoryear {%
Ahmed%
, Maume-Deschamps%
, Ribereau%
\BCBL {}\ \BBA {} Vial%
}{%
Ahmed%
\ \protect \BOthers {.}}{%
{\protect \APACyear {2017}}%
}]{%
ahmed2017semi}
\APACinsertmetastar {%
ahmed2017semi}%
\begin{APACrefauthors}%
Ahmed, M.%
, Maume-Deschamps, V.%
, Ribereau, P.%
\BCBL {}\ \BBA {} Vial, C.%
\end{APACrefauthors}%
\unskip\
\newblock
\APACrefYearMonthDay{2017}{}{}.
\newblock
{\BBOQ}\APACrefatitle {A semi-parametric estimation for max-mixture spatial
  processes} {A semi-parametric estimation for max-mixture spatial
  processes}.{\BBCQ}
\newblock
\APACjournalVolNumPages{arXiv preprint arXiv:1710.08120}{}{}{}.
\PrintBackRefs{\CurrentBib}

\bibitem [\protect \citeauthoryear {%
Alex%
, Sutskever%
\BCBL {}\ \BBA {} Hinton%
}{%
Alex%
\ \protect \BOthers {.}}{%
{\protect \APACyear {2012}}%
}]{%
NIPS2012_4824}
\APACinsertmetastar {%
NIPS2012_4824}%
\begin{APACrefauthors}%
Alex, K.%
, Sutskever, L.%
\BCBL {}\ \BBA {} Hinton, G.%
\end{APACrefauthors}%
\unskip\
\newblock
\APACrefYearMonthDay{2012}{}{}.
\newblock
{\BBOQ}\APACrefatitle {ImageNet Classification with Deep Convolutional Neural
  Networks} {Imagenet classification with deep convolutional neural
  networks}.{\BBCQ}
\newblock
\BIn{} F.~Pereira, C\BPBI J\BPBI C.~Burges, L.~Bottou\BCBL {}\ \BBA {} K\BPBI
  Q.~Weinberger\ (\BEDS), \APACrefbtitle {Advances in Neural Information
  Processing Systems 25} {Advances in neural information processing systems
  25}\ (\BPGS\ 1097--1105).
\newblock
\APACaddressPublisher{}{Curran Associates, Inc.}
\newblock
\begin{APACrefURL} \url{http://papers.nips.cc/paper/4824-imagenet-\\
  classification-with-deep-convolutional-neural-\\ networks.pdf}
  \end{APACrefURL}
\PrintBackRefs{\CurrentBib}

\bibitem [\protect \citeauthoryear {%
Bacro%
, Gaetan%
\BCBL {}\ \BBA {} Toulemonde%
}{%
Bacro%
\ \protect \BOthers {.}}{%
{\protect \APACyear {2016}}%
}]{%
bacro2016flexible}
\APACinsertmetastar {%
bacro2016flexible}%
\begin{APACrefauthors}%
Bacro, J.%
, Gaetan, C.%
\BCBL {}\ \BBA {} Toulemonde, G.%
\end{APACrefauthors}%
\unskip\
\newblock
\APACrefYearMonthDay{2016}{}{}.
\newblock
{\BBOQ}\APACrefatitle {A flexible dependence model for spatial extremes} {A
  flexible dependence model for spatial extremes}.{\BBCQ}
\newblock
\APACjournalVolNumPages{Journal of Statistical Planning and
  Inference}{172}{}{36--52}.
\PrintBackRefs{\CurrentBib}

\bibitem [\protect \citeauthoryear {%
Bortot%
, Coles%
\BCBL {}\ \BBA {} Tawn%
}{%
Bortot%
\ \protect \BOthers {.}}{%
{\protect \APACyear {2000}}%
}]{%
bortot2000multivariate}
\APACinsertmetastar {%
bortot2000multivariate}%
\begin{APACrefauthors}%
Bortot, P.%
, Coles, S.%
\BCBL {}\ \BBA {} Tawn, J.%
\end{APACrefauthors}%
\unskip\
\newblock
\APACrefYearMonthDay{2000}{}{}.
\newblock
{\BBOQ}\APACrefatitle {The multivariate gaussian tail model: An application to
  oceanographic data} {The multivariate gaussian tail model: An application to
  oceanographic data}.{\BBCQ}
\newblock
\APACjournalVolNumPages{Journal of the Royal Statistical Society: Series C
  (Applied Statistics)}{49}{1}{31--049}.
\PrintBackRefs{\CurrentBib}

\bibitem [\protect \citeauthoryear {%
Brown%
\ \BBA {} Resnick%
}{%
Brown%
\ \BBA {} Resnick%
}{%
{\protect \APACyear {1977}}%
}]{%
brown1977extreme}
\APACinsertmetastar {%
brown1977extreme}%
\begin{APACrefauthors}%
Brown, B\BPBI M.%
\BCBT {}\ \BBA {} Resnick, S\BPBI I.%
\end{APACrefauthors}%
\unskip\
\newblock
\APACrefYearMonthDay{1977}{}{}.
\newblock
{\BBOQ}\APACrefatitle {Extreme values of independent stochastic processes}
  {Extreme values of independent stochastic processes}.{\BBCQ}
\newblock
\APACjournalVolNumPages{Journal of Applied Probability}{14}{4}{732--739}.
\PrintBackRefs{\CurrentBib}

\bibitem [\protect \citeauthoryear {%
Caterini%
\ \BBA {} Chang%
}{%
Caterini%
\ \BBA {} Chang%
}{%
{\protect \APACyear {2018}}%
}]{%
caterini2018deep}
\APACinsertmetastar {%
caterini2018deep}%
\begin{APACrefauthors}%
Caterini, A\BPBI L.%
\BCBT {}\ \BBA {} Chang, D\BPBI E.%
\end{APACrefauthors}%
\unskip\
\newblock
\APACrefYear{2018}.
\newblock
\APACrefbtitle {Deep Neural Networks in a Mathematical Framework} {Deep neural
  networks in a mathematical framework}.
\newblock
\APACaddressPublisher{}{Springer}.
\PrintBackRefs{\CurrentBib}

\bibitem [\protect \citeauthoryear {%
Coles%
, Heffernan%
\BCBL {}\ \BBA {} Tawn%
}{%
Coles%
\ \protect \BOthers {.}}{%
{\protect \APACyear {1999}}%
}]{%
coles1999dependence}
\APACinsertmetastar {%
coles1999dependence}%
\begin{APACrefauthors}%
Coles, S.%
, Heffernan, J.%
\BCBL {}\ \BBA {} Tawn, J.%
\end{APACrefauthors}%
\unskip\
\newblock
\APACrefYearMonthDay{1999}{}{}.
\newblock
{\BBOQ}\APACrefatitle {Dependence measures for extreme value analyses}
  {Dependence measures for extreme value analyses}.{\BBCQ}
\newblock
\APACjournalVolNumPages{Extremes}{2}{4}{339--365}.
\PrintBackRefs{\CurrentBib}

\bibitem [\protect \citeauthoryear {%
Coles%
\ \BBA {} Pauli%
}{%
Coles%
\ \BBA {} Pauli%
}{%
{\protect \APACyear {2002}}%
}]{%
coles2002models}
\APACinsertmetastar {%
coles2002models}%
\begin{APACrefauthors}%
Coles, S.%
\BCBT {}\ \BBA {} Pauli, F.%
\end{APACrefauthors}%
\unskip\
\newblock
\APACrefYearMonthDay{2002}{}{}.
\newblock
{\BBOQ}\APACrefatitle {Models and inference for uncertainty in extremal
  dependence} {Models and inference for uncertainty in extremal
  dependence}.{\BBCQ}
\newblock
\APACjournalVolNumPages{Biometrika}{89}{1}{183--196}.
\PrintBackRefs{\CurrentBib}

\bibitem [\protect \citeauthoryear {%
De~Haan%
\ \BBA {} Ferreira%
}{%
De~Haan%
\ \BBA {} Ferreira%
}{%
{\protect \APACyear {2007}}%
}]{%
de2007extreme}
\APACinsertmetastar {%
de2007extreme}%
\begin{APACrefauthors}%
De~Haan, L.%
\BCBT {}\ \BBA {} Ferreira, A.%
\end{APACrefauthors}%
\unskip\
\newblock
\APACrefYear{2007}.
\newblock
\APACrefbtitle {Extreme value theory: an introduction} {Extreme value theory:
  an introduction}.
\newblock
\APACaddressPublisher{}{Springer Science \& Business Media}.
\PrintBackRefs{\CurrentBib}

\bibitem [\protect \citeauthoryear {%
De~Haan%
\ \protect \BOthers {.}}{%
De~Haan%
\ \protect \BOthers {.}}{%
{\protect \APACyear {1984}}%
}]{%
de1984spectral}
\APACinsertmetastar {%
de1984spectral}%
\begin{APACrefauthors}%
De~Haan, L.%
\BCBT {}\ \BOthersPeriod {.}
\end{APACrefauthors}%
\unskip\
\newblock
\APACrefYearMonthDay{1984}{}{}.
\newblock
{\BBOQ}\APACrefatitle {A spectral representation for max-stable processes} {A
  spectral representation for max-stable processes}.{\BBCQ}
\newblock
\APACjournalVolNumPages{The annals of probability}{12}{4}{1194--1204}.
\PrintBackRefs{\CurrentBib}

\bibitem [\protect \citeauthoryear {%
De~Haan%
, Pereira%
\BCBL {}\ \protect \BOthers {.}}{%
De~Haan%
\ \protect \BOthers {.}}{%
{\protect \APACyear {2006}}%
}]{%
de2006spatial}
\APACinsertmetastar {%
de2006spatial}%
\begin{APACrefauthors}%
De~Haan, L.%
, Pereira, T\BPBI T.%
\BCBL {}\ \BOthersPeriod {.}\end{APACrefauthors}%
\unskip\
\newblock
\APACrefYearMonthDay{2006}{}{}.
\newblock
{\BBOQ}\APACrefatitle {Spatial extremes: Models for the stationary case}
  {Spatial extremes: Models for the stationary case}.{\BBCQ}
\newblock
\APACjournalVolNumPages{The annals of statistics}{34}{1}{146--168}.
\PrintBackRefs{\CurrentBib}

\bibitem [\protect \citeauthoryear {%
Embrechts%
, Kl{\"u}ppelberg%
\BCBL {}\ \BBA {} Mikosch%
}{%
Embrechts%
\ \protect \BOthers {.}}{%
{\protect \APACyear {2013}}%
}]{%
embrechts2013modelling}
\APACinsertmetastar {%
embrechts2013modelling}%
\begin{APACrefauthors}%
Embrechts, P.%
, Kl{\"u}ppelberg, C.%
\BCBL {}\ \BBA {} Mikosch, T.%
\end{APACrefauthors}%
\unskip\
\newblock
\APACrefYear{2013}.
\newblock
\APACrefbtitle {Modelling extremal events: for insurance and finance}
  {Modelling extremal events: for insurance and finance}\ (\BVOL~33).
\newblock
\APACaddressPublisher{}{Springer Science \& Business Media}.
\PrintBackRefs{\CurrentBib}

\bibitem [\protect \citeauthoryear {%
Fukushima%
}{%
Fukushima%
}{%
{\protect \APACyear {1980}}%
}]{%
fukushima1980neocognitron}
\APACinsertmetastar {%
fukushima1980neocognitron}%
\begin{APACrefauthors}%
Fukushima, K.%
\end{APACrefauthors}%
\unskip\
\newblock
\APACrefYearMonthDay{1980}{}{}.
\newblock
{\BBOQ}\APACrefatitle {Neocognitron: A self-organizing neural network model for
  a mechanism of pattern recognition unaffected by shift in position}
  {Neocognitron: A self-organizing neural network model for a mechanism of
  pattern recognition unaffected by shift in position}.{\BBCQ}
\newblock
\APACjournalVolNumPages{Biological cybernetics}{36}{4}{193--202}.
\PrintBackRefs{\CurrentBib}

\bibitem [\protect \citeauthoryear {%
He%
, Zhang%
, Ren%
\BCBL {}\ \BBA {} Sun%
}{%
He%
\ \protect \BOthers {.}}{%
{\protect \APACyear {2016}}%
}]{%
he2016deep}
\APACinsertmetastar {%
he2016deep}%
\begin{APACrefauthors}%
He, K.%
, Zhang, X.%
, Ren, S.%
\BCBL {}\ \BBA {} Sun, J.%
\end{APACrefauthors}%
\unskip\
\newblock
\APACrefYearMonthDay{2016}{}{}.
\newblock
{\BBOQ}\APACrefatitle {Deep residual learning for image recognition} {Deep
  residual learning for image recognition}.{\BBCQ}
\newblock
\BIn{} \APACrefbtitle {Proceedings of the IEEE conference on computer vision
  and pattern recognition} {Proceedings of the ieee conference on computer
  vision and pattern recognition}\ (\BPGS\ 770--778).
\PrintBackRefs{\CurrentBib}

\bibitem [\protect \citeauthoryear {%
Hooker%
, Duveiller%
\BCBL {}\ \BBA {} Cescatti%
}{%
Hooker%
\ \protect \BOthers {.}}{%
{\protect \APACyear {2018}}%
}]{%
hooker2018global}
\APACinsertmetastar {%
hooker2018global}%
\begin{APACrefauthors}%
Hooker, J.%
, Duveiller, G.%
\BCBL {}\ \BBA {} Cescatti, A.%
\end{APACrefauthors}%
\unskip\
\newblock
\APACrefYearMonthDay{2018}{}{}.
\newblock
{\BBOQ}\APACrefatitle {A global dataset of air temperature derived from
  satellite remote sensing and weather stations} {A global dataset of air
  temperature derived from satellite remote sensing and weather
  stations}.{\BBCQ}
\newblock
\APACjournalVolNumPages{Scientific data}{5}{1}{1--11}.
\PrintBackRefs{\CurrentBib}

\bibitem [\protect \citeauthoryear {%
Hubel%
\ \BBA {} Wiesel%
}{%
Hubel%
\ \BBA {} Wiesel%
}{%
{\protect \APACyear {1968}}%
}]{%
hubel1968receptive}
\APACinsertmetastar {%
hubel1968receptive}%
\begin{APACrefauthors}%
Hubel, D\BPBI H.%
\BCBT {}\ \BBA {} Wiesel, T\BPBI N.%
\end{APACrefauthors}%
\unskip\
\newblock
\APACrefYearMonthDay{1968}{}{}.
\newblock
{\BBOQ}\APACrefatitle {Receptive fields and functional architecture of monkey
  striate cortex} {Receptive fields and functional architecture of monkey
  striate cortex}.{\BBCQ}
\newblock
\APACjournalVolNumPages{The Journal of physiology}{195}{1}{215--243}.
\PrintBackRefs{\CurrentBib}

\bibitem [\protect \citeauthoryear {%
Huser%
}{%
Huser%
}{%
{\protect \APACyear {2020}}%
}]{%
huser2020eva}
\APACinsertmetastar {%
huser2020eva}%
\begin{APACrefauthors}%
Huser, R.%
\end{APACrefauthors}%
\unskip\
\newblock
\APACrefYearMonthDay{2020}{}{}.
\newblock
{\BBOQ}\APACrefatitle {EVA 2019 data competition on spatio-temporal prediction
  of Red Sea surface temperature extremes} {Eva 2019 data competition on
  spatio-temporal prediction of red sea surface temperature extremes}.{\BBCQ}
\newblock
\APACjournalVolNumPages{Extremes}{}{}{1--14}.
\PrintBackRefs{\CurrentBib}

\bibitem [\protect \citeauthoryear {%
Kabluchko%
, Schlather%
, De~Haan%
\BCBL {}\ \protect \BOthers {.}}{%
Kabluchko%
\ \protect \BOthers {.}}{%
{\protect \APACyear {2009}}%
}]{%
kabluchko2009stationary}
\APACinsertmetastar {%
kabluchko2009stationary}%
\begin{APACrefauthors}%
Kabluchko, Z.%
, Schlather, M.%
, De~Haan, L.%
\BCBL {}\ \BOthersPeriod {.}\end{APACrefauthors}%
\unskip\
\newblock
\APACrefYearMonthDay{2009}{}{}.
\newblock
{\BBOQ}\APACrefatitle {Stationary max-stable fields associated to negative
  definite functions} {Stationary max-stable fields associated to negative
  definite functions}.{\BBCQ}
\newblock
\APACjournalVolNumPages{The Annals of Probability}{37}{5}{2042--2065}.
\PrintBackRefs{\CurrentBib}

\bibitem [\protect \citeauthoryear {%
Kingma%
\ \BBA {} Ba%
}{%
Kingma%
\ \BBA {} Ba%
}{%
{\protect \APACyear {2014}}%
}]{%
kingma2014adam}
\APACinsertmetastar {%
kingma2014adam}%
\begin{APACrefauthors}%
Kingma, D.%
\BCBT {}\ \BBA {} Ba, J.%
\end{APACrefauthors}%
\unskip\
\newblock
\APACrefYearMonthDay{2014}{}{}.
\newblock
{\BBOQ}\APACrefatitle {Adam: A method for stochastic optimization} {Adam: A
  method for stochastic optimization}.{\BBCQ}
\newblock
\APACjournalVolNumPages{arXiv preprint arXiv:1412.6980}{}{}{}.
\PrintBackRefs{\CurrentBib}

\bibitem [\protect \citeauthoryear {%
LeCun%
\ \protect \BOthers {.}}{%
LeCun%
\ \protect \BOthers {.}}{%
{\protect \APACyear {1990}}%
}]{%
lecun1990handwritten}
\APACinsertmetastar {%
lecun1990handwritten}%
\begin{APACrefauthors}%
LeCun, Y.%
, Boser, B\BPBI E.%
, Denker, J\BPBI S.%
, Henderson, D.%
, Howard, R\BPBI E.%
, Hubbard, W\BPBI E.%
\BCBL {}\ \BBA {} Jackel, L\BPBI D.%
\end{APACrefauthors}%
\unskip\
\newblock
\APACrefYearMonthDay{1990}{}{}.
\newblock
{\BBOQ}\APACrefatitle {Handwritten digit recognition with a back-propagation
  network} {Handwritten digit recognition with a back-propagation
  network}.{\BBCQ}
\newblock
\BIn{} \APACrefbtitle {Advances in neural information processing systems}
  {Advances in neural information processing systems}\ (\BPGS\ 396--404).
\PrintBackRefs{\CurrentBib}

\bibitem [\protect \citeauthoryear {%
Ledford%
\ \BBA {} Tawn%
}{%
Ledford%
\ \BBA {} Tawn%
}{%
{\protect \APACyear {1996}}%
}]{%
ledford1996statistics}
\APACinsertmetastar {%
ledford1996statistics}%
\begin{APACrefauthors}%
Ledford, A.%
\BCBT {}\ \BBA {} Tawn, J\BPBI A.%
\end{APACrefauthors}%
\unskip\
\newblock
\APACrefYearMonthDay{1996}{}{}.
\newblock
{\BBOQ}\APACrefatitle {Statistics for near independence in multivariate extreme
  values} {Statistics for near independence in multivariate extreme
  values}.{\BBCQ}
\newblock
\APACjournalVolNumPages{Biometrika}{83}{1}{169--187}.
\PrintBackRefs{\CurrentBib}

\bibitem [\protect \citeauthoryear {%
Lin%
\ \protect \BOthers {.}}{%
Lin%
\ \protect \BOthers {.}}{%
{\protect \APACyear {2018}}%
}]{%
lin2018exploiting}
\APACinsertmetastar {%
lin2018exploiting}%
\begin{APACrefauthors}%
Lin, Y.%
, Mago, N.%
, Gao, Y.%
, Li, Y.%
, Chiang, Y.%
, Shahabi, C.%
\BCBL {}\ \BBA {} Ambite, J\BPBI L.%
\end{APACrefauthors}%
\unskip\
\newblock
\APACrefYearMonthDay{2018}{}{}.
\newblock
{\BBOQ}\APACrefatitle {Exploiting spatiotemporal patterns for accurate air
  quality forecasting using deep learning} {Exploiting spatiotemporal patterns
  for accurate air quality forecasting using deep learning}.{\BBCQ}
\newblock
\BIn{} \APACrefbtitle {Proceedings of the 26th ACM SIGSPATIAL International
  Conference on Advances in Geographic Information Systems} {Proceedings of the
  26th acm sigspatial international conference on advances in geographic
  information systems}\ (\BPGS\ 359--368).
\PrintBackRefs{\CurrentBib}

\bibitem [\protect \citeauthoryear {%
Liu%
\ \protect \BOthers {.}}{%
Liu%
\ \protect \BOthers {.}}{%
{\protect \APACyear {2016}}%
}]{%
liu2016application}
\APACinsertmetastar {%
liu2016application}%
\begin{APACrefauthors}%
Liu, Y.%
, Racah, E.%
, Correa, J.%
, Khosrowshahi, A.%
, Lavers, D.%
, Kunkel, K.%
\BDBL {}others%
\end{APACrefauthors}%
\unskip\
\newblock
\APACrefYearMonthDay{2016}{}{}.
\newblock
{\BBOQ}\APACrefatitle {Application of deep convolutional neural networks for
  detecting extreme weather in climate datasets} {Application of deep
  convolutional neural networks for detecting extreme weather in climate
  datasets}.{\BBCQ}
\newblock
\APACjournalVolNumPages{arXiv preprint arXiv:1605.01156}{}{}{}.
\PrintBackRefs{\CurrentBib}

\bibitem [\protect \citeauthoryear {%
Opitz%
}{%
Opitz%
}{%
{\protect \APACyear {2013}}%
}]{%
opitz2013extremal}
\APACinsertmetastar {%
opitz2013extremal}%
\begin{APACrefauthors}%
Opitz, T.%
\end{APACrefauthors}%
\unskip\
\newblock
\APACrefYearMonthDay{2013}{}{}.
\newblock
{\BBOQ}\APACrefatitle {Extremal t processes: Elliptical domain of attraction
  and a spectral representation} {Extremal t processes: Elliptical domain of
  attraction and a spectral representation}.{\BBCQ}
\newblock
\APACjournalVolNumPages{Journal of Multivariate Analysis}{122}{}{409--413}.
\PrintBackRefs{\CurrentBib}

\bibitem [\protect \citeauthoryear {%
Schlather%
}{%
Schlather%
}{%
{\protect \APACyear {2002}}%
}]{%
schlather2002models}
\APACinsertmetastar {%
schlather2002models}%
\begin{APACrefauthors}%
Schlather, M.%
\end{APACrefauthors}%
\unskip\
\newblock
\APACrefYearMonthDay{2002}{}{}.
\newblock
{\BBOQ}\APACrefatitle {Models for stationary max-stable random fields} {Models
  for stationary max-stable random fields}.{\BBCQ}
\newblock
\APACjournalVolNumPages{Extremes}{5}{1}{33--44}.
\PrintBackRefs{\CurrentBib}

\bibitem [\protect \citeauthoryear {%
Schlather%
\ \BBA {} Tawn%
}{%
Schlather%
\ \BBA {} Tawn%
}{%
{\protect \APACyear {2003}}%
}]{%
schlather2003dependence}
\APACinsertmetastar {%
schlather2003dependence}%
\begin{APACrefauthors}%
Schlather, M.%
\BCBT {}\ \BBA {} Tawn, J\BPBI A.%
\end{APACrefauthors}%
\unskip\
\newblock
\APACrefYearMonthDay{2003}{}{}.
\newblock
{\BBOQ}\APACrefatitle {A dependence measure for multivariate and spatial
  extreme values: Properties and inference} {A dependence measure for
  multivariate and spatial extreme values: Properties and inference}.{\BBCQ}
\newblock
\APACjournalVolNumPages{Biometrika}{90}{1}{139--156}.
\PrintBackRefs{\CurrentBib}

\bibitem [\protect \citeauthoryear {%
Smith%
}{%
Smith%
}{%
{\protect \APACyear {1990}}%
}]{%
smith1990max}
\APACinsertmetastar {%
smith1990max}%
\begin{APACrefauthors}%
Smith, R\BPBI L.%
\end{APACrefauthors}%
\unskip\
\newblock
\APACrefYearMonthDay{1990}{}{}.
\newblock
{\BBOQ}\APACrefatitle {Max-stable processes and spatial extremes} {Max-stable
  processes and spatial extremes}.{\BBCQ}
\newblock
\APACjournalVolNumPages{Unpublished manuscript}{205}{}{}.
\PrintBackRefs{\CurrentBib}

\bibitem [\protect \citeauthoryear {%
Wadsworth%
\ \BBA {} Tawn%
}{%
Wadsworth%
\ \BBA {} Tawn%
}{%
{\protect \APACyear {2012}}%
}]{%
wadsworth2012dependence}
\APACinsertmetastar {%
wadsworth2012dependence}%
\begin{APACrefauthors}%
Wadsworth, J\BPBI L.%
\BCBT {}\ \BBA {} Tawn, J\BPBI A.%
\end{APACrefauthors}%
\unskip\
\newblock
\APACrefYearMonthDay{2012}{}{}.
\newblock
{\BBOQ}\APACrefatitle {Dependence modelling for spatial extremes} {Dependence
  modelling for spatial extremes}.{\BBCQ}
\newblock
\APACjournalVolNumPages{Biometrika}{99}{2}{253--272}.
\PrintBackRefs{\CurrentBib}

\bibitem [\protect \citeauthoryear {%
Wang%
, Cao%
\BCBL {}\ \BBA {} Yu%
}{%
Wang%
\ \protect \BOthers {.}}{%
{\protect \APACyear {2019}}%
}]{%
wang2019deep}
\APACinsertmetastar {%
wang2019deep}%
\begin{APACrefauthors}%
Wang, S.%
, Cao, J.%
\BCBL {}\ \BBA {} Yu, P\BPBI S.%
\end{APACrefauthors}%
\unskip\
\newblock
\APACrefYearMonthDay{2019}{}{}.
\newblock
{\BBOQ}\APACrefatitle {Deep learning for spatio-temporal data mining: A survey}
  {Deep learning for spatio-temporal data mining: A survey}.{\BBCQ}
\newblock
\APACjournalVolNumPages{arXiv preprint arXiv:1906.04928}{}{}{}.
\PrintBackRefs{\CurrentBib}

\bibitem [\protect \citeauthoryear {%
Xie%
, Girshick%
, Doll{\'a}r%
, Tu%
\BCBL {}\ \BBA {} He%
}{%
Xie%
\ \protect \BOthers {.}}{%
{\protect \APACyear {2017}}%
}]{%
xie2017aggregated}
\APACinsertmetastar {%
xie2017aggregated}%
\begin{APACrefauthors}%
Xie, S.%
, Girshick, R.%
, Doll{\'a}r, P.%
, Tu, Z.%
\BCBL {}\ \BBA {} He, K.%
\end{APACrefauthors}%
\unskip\
\newblock
\APACrefYearMonthDay{2017}{}{}.
\newblock
{\BBOQ}\APACrefatitle {Aggregated residual transformations for deep neural
  networks} {Aggregated residual transformations for deep neural
  networks}.{\BBCQ}
\newblock
\BIn{} \APACrefbtitle {Proceedings of the IEEE conference on computer vision
  and pattern recognition} {Proceedings of the ieee conference on computer
  vision and pattern recognition}\ (\BPGS\ 1492--1500).
\PrintBackRefs{\CurrentBib}

\bibitem [\protect \citeauthoryear {%
Yamashita%
, Nishio%
, Do%
\BCBL {}\ \BBA {} Togashi%
}{%
Yamashita%
\ \protect \BOthers {.}}{%
{\protect \APACyear {2018}}%
}]{%
yamashita2018convolutional}
\APACinsertmetastar {%
yamashita2018convolutional}%
\begin{APACrefauthors}%
Yamashita, R.%
, Nishio, M.%
, Do, R.%
\BCBL {}\ \BBA {} Togashi, K.%
\end{APACrefauthors}%
\unskip\
\newblock
\APACrefYearMonthDay{2018}{}{}.
\newblock
{\BBOQ}\APACrefatitle {Convolutional neural networks: an overview and
  application in radiology} {Convolutional neural networks: an overview and
  application in radiology}.{\BBCQ}
\newblock
\APACjournalVolNumPages{Insights into imaging}{9}{4}{611--629}.
\PrintBackRefs{\CurrentBib}

\bibitem [\protect \citeauthoryear {%
Yu%
, Uy%
\BCBL {}\ \BBA {} Dauwels%
}{%
Yu%
\ \protect \BOthers {.}}{%
{\protect \APACyear {2016}}%
}]{%
yu2016modeling}
\APACinsertmetastar {%
yu2016modeling}%
\begin{APACrefauthors}%
Yu, H.%
, Uy, W\BPBI I\BPBI T.%
\BCBL {}\ \BBA {} Dauwels, J.%
\end{APACrefauthors}%
\unskip\
\newblock
\APACrefYearMonthDay{2016}{}{}.
\newblock
{\BBOQ}\APACrefatitle {Modeling spatial extremes via ensemble-of-trees of
  pairwise copulas} {Modeling spatial extremes via ensemble-of-trees of
  pairwise copulas}.{\BBCQ}
\newblock
\APACjournalVolNumPages{IEEE Transactions on Signal
  Processing}{65}{3}{571--586}.
\PrintBackRefs{\CurrentBib}

\bibitem [\protect \citeauthoryear {%
Zhu%
, Chen%
, Zhu%
, Duan%
\BCBL {}\ \BBA {} Liu%
}{%
Zhu%
\ \protect \BOthers {.}}{%
{\protect \APACyear {2018}}%
}]{%
zhu2018wind}
\APACinsertmetastar {%
zhu2018wind}%
\begin{APACrefauthors}%
Zhu, Q.%
, Chen, J.%
, Zhu, L.%
, Duan, X.%
\BCBL {}\ \BBA {} Liu, Y.%
\end{APACrefauthors}%
\unskip\
\newblock
\APACrefYearMonthDay{2018}{}{}.
\newblock
{\BBOQ}\APACrefatitle {Wind speed prediction with spatio--temporal correlation:
  A deep learning approach} {Wind speed prediction with spatio--temporal
  correlation: A deep learning approach}.{\BBCQ}
\newblock
\APACjournalVolNumPages{Energies}{11}{4}{705}.
\PrintBackRefs{\CurrentBib}

\bibitem [\protect \citeauthoryear {%
Zuo%
\ \protect \BOthers {.}}{%
Zuo%
\ \protect \BOthers {.}}{%
{\protect \APACyear {2015}}%
}]{%
zuo2015convolutional}
\APACinsertmetastar {%
zuo2015convolutional}%
\begin{APACrefauthors}%
Zuo, Z.%
, Shuai, B.%
, Wang, G.%
, Liu, X.%
, Wang, X.%
, Wang, B.%
\BCBL {}\ \BBA {} Chen, Y.%
\end{APACrefauthors}%
\unskip\
\newblock
\APACrefYearMonthDay{2015}{}{}.
\newblock
{\BBOQ}\APACrefatitle {Convolutional recurrent neural networks: Learning
  spatial dependencies for image representation} {Convolutional recurrent
  neural networks: Learning spatial dependencies for image
  representation}.{\BBCQ}
\newblock
\BIn{} \APACrefbtitle {Proceedings of the IEEE conference on computer vision
  and pattern recognition workshops} {Proceedings of the ieee conference on
  computer vision and pattern recognition workshops}\ (\BPGS\ 18--26).
\PrintBackRefs{\CurrentBib}

\end{thebibliography}

\end{document}